\documentclass[%
 aip,
 amsmath,amssymb,
preprint,%
]{revtex4-1}

\usepackage{graphicx}
\graphicspath{{./Figures/}} 
\usepackage{dcolumn}
\usepackage{bm}
\usepackage{color}

\usepackage[utf8]{inputenc}
\usepackage[T1]{fontenc}
\usepackage{mathptmx}
\usepackage{chemarrow}
\usepackage{etoolbox}
\usepackage[version=4]{mhchem}        
\usepackage{float}
\usepackage{makecell}
\usepackage{enumitem}
\usepackage{amsfonts}
\usepackage{soul}                   

\makeatletter
\def\@email#1#2{%
 \endgroup
 \patchcmd{\titleblock@produce}
  {\frontmatter@RRAPformat}
  {\frontmatter@RRAPformat{\produce@RRAP{*#1\href{mailto:#2}{#2}}}\frontmatter@RRAPformat}
  {}{}
}
\makeatletter
\AtBeginDocument{\let\hl\@firstofone}
\makeatother

\makeatother
\begin{document}

\preprint{arXiv version}

\title{Possible Origin of Preformed Hole Pairs and Superconductivity in Cuprates}

\author{Shu Wang}
\affiliation{Department of Materials Science and Engineering, University of California Berkeley, Berkeley, CA 94720, USA}

\author{Joel W. Ager}
\email{jwager@lbl.gov}
\affiliation{Department of Materials Science and Engineering, University of California Berkeley, Berkeley, CA 94720, USA}
\affiliation{ Materials Sciences Division, Lawrence Berkeley National Laboratory, Berkeley, CA 94720, USA}
\affiliation{Berkeley Educational Alliance for Research in Singapore (BEARS), Ltd., 1 CREATE Way, 138602, Singapore}

\author{Wladek Walukiewicz}
\email{w\_walukiewicz@lbl.gov}
\affiliation{ Materials Sciences Division, Lawrence Berkeley National Laboratory, Berkeley, CA 94720, USA}
\affiliation{Berkeley Educational Alliance for Research in Singapore (BEARS), Ltd., 1 CREATE Way, 138602, Singapore}
\affiliation{Saule Research Institute, Wroclaw Technology Park, 11 Dunska Street, Sigma Building, 54-130 Wrocław, Poland}

\date{\today}

\begin{abstract}
This paper addresses the long standing and controversial issue of the origin of superconductivity in cuprates. It is shown that the superconductivity can be attributed to amphoteric defects associated with vacancy sites in copper oxide planes. 
A local defect lattice relaxation results in a negative-$U$ energy binding two holes on amphoteric defects in the donor configuration that act as preformed boson pair. Thermodynamic equilibrium between defects in the donor and acceptor
configurations stabilizes Fermi energy at the amphoteric defect charge transition state assuring resonant coupling between free holes and the localized hole pairs. Model calculations provide explanation for most important superconducting properties of cuprates. They show that the critical temperature is primarily determined by the density of the amphoteric defects in the donor configuration. This explains ubiquity of dome-like dependence of the critical temperature on the doping as well as its universal dependence on the superfluid density. Intentional doping with chemical acceptors or donors is neither necessary nor sufficient condition for superconductivity that is fully determined by the amphoteric defects whose concentration can be controlled by crystal nonstoichiometry. The only role of chemical doping is changing the balance between concentrations of amphoteric defects in the donor and acceptor configurations resulting in an increase of the superfluid density and thus also the critical
temperature for acceptor and a decrease for donor doping. This accounts for the experimentally observed distinct asymmetry between the dome structures for the chemical doping with acceptors and donors. The unusual sensitivity of the critical temperature to external perturbations is explained by the resonant nature of the coupling between free holes and preformed hole pairs. The work has broader implications as it could be applicable to other superconductors with dome-like dependence of the
critical temperature on doping.
\end{abstract}

\maketitle
\tableofcontents
\newpage

\section{Introduction}

The discovery of high temperature superconductivity in 
cuprates (cupreous oxides) \cite{Bednorz1986} 
was immediately followed by explosive
research that unearthed a large number of materials with critical
temperatures as high as 133 K at ambient and even
higher at high hydrostatic pressures.\cite{Schilling1993, Chu2015} This
unprecedented triumph of empirical material science created a difficult
challenge for existing theories of superconductivity and despite of more
than 35 years of research there is still no generally accepted
explanation of the origin of the superconductivity in these materials.
Although there were some indications of possible role of lattice
vibrations \cite{Lee2006a}, an absence of consistent isotope shift
strongly indicated an electronic origin of the superconductivity.\cite{Zhao2001} 
Initially a considerable attention was devoted to the model of resonating valence band (RVB) \cite{Anderson1959,Anderson1987} that
was able to address some of unique features of the normal state of
cuprates. Separately it was suggested that the hole pairing and
superconductivity could originate from antiferromagnetic spin
fluctuations.\cite{Chubukov1997} Also, several research groups have
pursued theories based on quantum critical point (QCP) concept where it
is argued that the properties of the material are determined by a hidden
competing order parameter.\cite{Simon2002}

Another broad class of models of superconductivity invoked the original
concept of preformed bosons in which superconductivity originates from
interaction of free charge carriers with preformed pairs of charges.\cite{Schafroth1954} 
Different variations of this concept were actively
pursued by several research groups and theoretical formalisms have been
developed to understand superconductivity in various unconventional
superconductors\cite{Simanek1979,Ting1980,Ting1980a} including superconducting
cuprates.\cite{Friedberg1989,Ranninger1988, Bar-Yam1991} The main difficulty with these approaches was that they could not identify a physical process for
formation of well-developed, robust negative-$U$ centers. In addition they
required for the Fermi energy to be in close resonance with the charge
transition state of the negative-$U$ centers.\cite{Ting1980} It was
not clear how these demanding conditions could be so readily satisfied
in such large variety of the high $T_c$ cuprates.

Over more than three decades the expansive experimental and theoretical
research on superconducting cuprates was summarized in many
comprehensive review articles which
demonstrate that there is a general consensus on several well-understood
key properties common to these materials.\cite{Chu2015, Norman2003,Pickett1989,Orenstein2000,Keimer2015} 
Thus, superconducting cuprates
crystalize in a basic perovskite-like structure. The parent undoped stoichiometric compounds are antiferromagnetic charge transfer insulators with a large energy gap between fully occupied charge
transfer band (CBT) and empty upper Hubbard band (UHB). Charge carriers
can be introduced and electrical properties can be changed by varying
crystal stoichiometry (Cu/O ratio) and by chemical doping of the charge
reservoir building blocks. The electrical properties of the cuprates are
highly anisotropic with the conductivity confined to 2-dimensional
\ce{CuO2} planes.

In contrast, there is no common explanation or understanding of a
variety of experimental observations related to the superconductivity:

\begin{enumerate}
\def\labelenumi{\arabic{enumi})}
\item
  Optimization of superconducting properties require material processing
  steps involving high temperature annealing and cooling under specific,
  well controlled oxygen ambient conditions.
\item
  The superconducting transition temperature shows a dome-like
  dependence on doping.
\item
  Intentional chemical p-type doping augments whereas n-type doping
  diminishes superconducting properties.
\item
  Critical temperatures for different cuprates show a universal
  dependence on the superconducting condensate or superfluid density.
\item
  Critical temperatures are strongly dependent on external
  perturbations: hydrostatic pressure and electrical bias.
\end{enumerate}

In this paper we show that superconductivity in cuprates can be well
described by an amphoteric defect model in which vacancy like defects in
\ce{CuO2} planes are negative-$U$ centers that can undergo a
transformation between acceptor and donor configuration. The defects in
the donor configuration act as localized hole pairs. The boson field of
the randomly distributed pairs induces hole superconductivity at low
temperature. The paper is organized in several sections that address
different aspects of the superconducting properties of cuprates. In
section II we briefly introduce the concept of negative-$U$ amphoteric
defects that has been originally developed for semiconducting materials.
An extension of the model to the case of superconducting cuprates with
amphoteric defects located in 2-dimensional \ce{CuO2} planes
is presented in section III. The key results of the paper are presented
in section IV where it is shown how coupling of free holes to pairs of
holes localized on amphoteric defects in the donor configuration leads
to high temperature superconductivity. It also shows how the increased
doping leads to a reduction of the preformed pairs and termination of
the critical temperature and high defect concentration providing a
straightforward explanation for the origin of the dome like dependence
of the critical temperature on the doping. The effects of the chemical
doping on the critical temperature are discussed in section V where it
is shown that in both cases the superconductivity is associated with
free holes in the partially occupied charge transfer band. Calculations
account very well for the asymmetry in the shape of the superconducting
dome observed in n-type and p-type doped cuprates. Section VI deals with
the effects of external perturbations such as hydrostatic pressure and
external electric field bias on the superconducting properties. Finally,
section VII summarizes the results of the paper and offers an outlook on
possible applications of the amphoteric defect model to other
unconventional superconductors.


\section{Amphoteric Defects and Localized negative-$U$ Centers}

There are several unique features of cuprates that distinguish them from
classical superconductors. Most notably these materials do not exhibit
standard metallic behavior but rather show semiconducting properties
where the conductivities and carrier concentrations can be modified and
controlled by intentional chemical doping with acceptors or donors
and/or by varying crystal stoichiometry. Also, in a striking similarity
to semiconductors whose properties are sensitive to external
perturbations properties of superconducting cuprates, including the
critical temperature were found to be very sensitive to hydrostatic
pressure\cite{Gao1994} and electrical bias of thin films.\cite{Bollinger2011,Leng2011} 
This is strongly suggestive that physical
concepts well developed and commonly used in semiconductors could
provide a guidance to understanding properties of cuprates.

Localized negative-$U$ centers i.e., centers with a negative effective
electrostatic interaction energy between two electrons or two holes are
well known and have been extensively studied in semiconductors. In most
instances the electrostatic repulsion is overcome by large attractive
potentials associated with local lattice relaxation effects. Negative-$U$
centers have been invoked in explanation of a variety of effects observed in
semiconductors.\cite{Watkins1980,Baraff1980,Coutinho2020} 
Notably properties of the extensively studied DX centers were explained by a negative-$U$ behavior of substitutional donors in GaAs.\cite{Chadi1988} Independently it
was shown that vacancy-like defects exhibit negative-$U$ behavior as they
undergo transformation between negatively charged acceptor-like and
positively charged donor-like configuration.\cite{Baraff1985} The
actual configuration depends on the Fermi energy and, for example, in
GaAs a negatively charged gallium vacancy V\textsubscript{Ga} acceptor
is stable in n-type material but relaxes to a positively charged
(V\textsubscript{As}+As\textsubscript{Ga}) donor complex configuration
in p-type material.\cite{Baraff1985} The transformation between
these two configurations occurs at the Fermi energy at which the
formation energy of a negatively charged acceptor equals the formation
energy of positively charged donor.

The amphoteric nature of dangling bond defects leads to the well-known
effect of the defect induced stabilization of the Fermi energy\cite{Walukiewicz1987,Walukiewicz1988,Walukiewicz1988a}
 where it was demonstrated that in semiconductors
the stabilization energy, $E$\textsubscript{FS} remains constant and is
located at about 4.9 eV below the vacuum level.\cite{Walukiewicz1988a} The
same phenomenon has been shown to account for the pinning of the Fermi
energy on semiconductor surfaces and at metal/semiconductor interfaces.\cite{Walukiewicz1987} 
Also it has provided a universal explanation for large
diversity in the doping properties of semiconductors.\cite{Walukiewicz2001} Most recently the amphoteric defect model was used
to understand the unique properties of hybrid organic-inorganic
perovskites and to develop a new mechanism of a dynamic doping and p/n
junction formation in these materials.\cite{Walukiewicz2018,Walukiewicz2020,Walukiewicz2021}

In a presence of vacancy-like native amphoteric defects, the electrical
properties of a semiconductor material are determined by the location of
the Fermi level stabilization energy $E$\textsubscript{FS} relative to the
edges of the bands of extended states.\cite{Walukiewicz1988a,Walukiewicz2001} In
standard semiconductors the valence band and the conduction band edges
originate from orbitals of the component atoms. This results in large
variations of the electron affinity and/or ionization energy and thus
also band offsets between different materials. Consequently, as is
illustrated in \textbf{Fig. 1}, 
this leads to large variations of the
locations of semiconductor band edges relative to the
$E$\textsubscript{FS} that remains constant at about 4.9 eV below the
vacuum level for all materials.

\begin{figure}
\centering
 \includegraphics[width=6in,height=3.87153in]{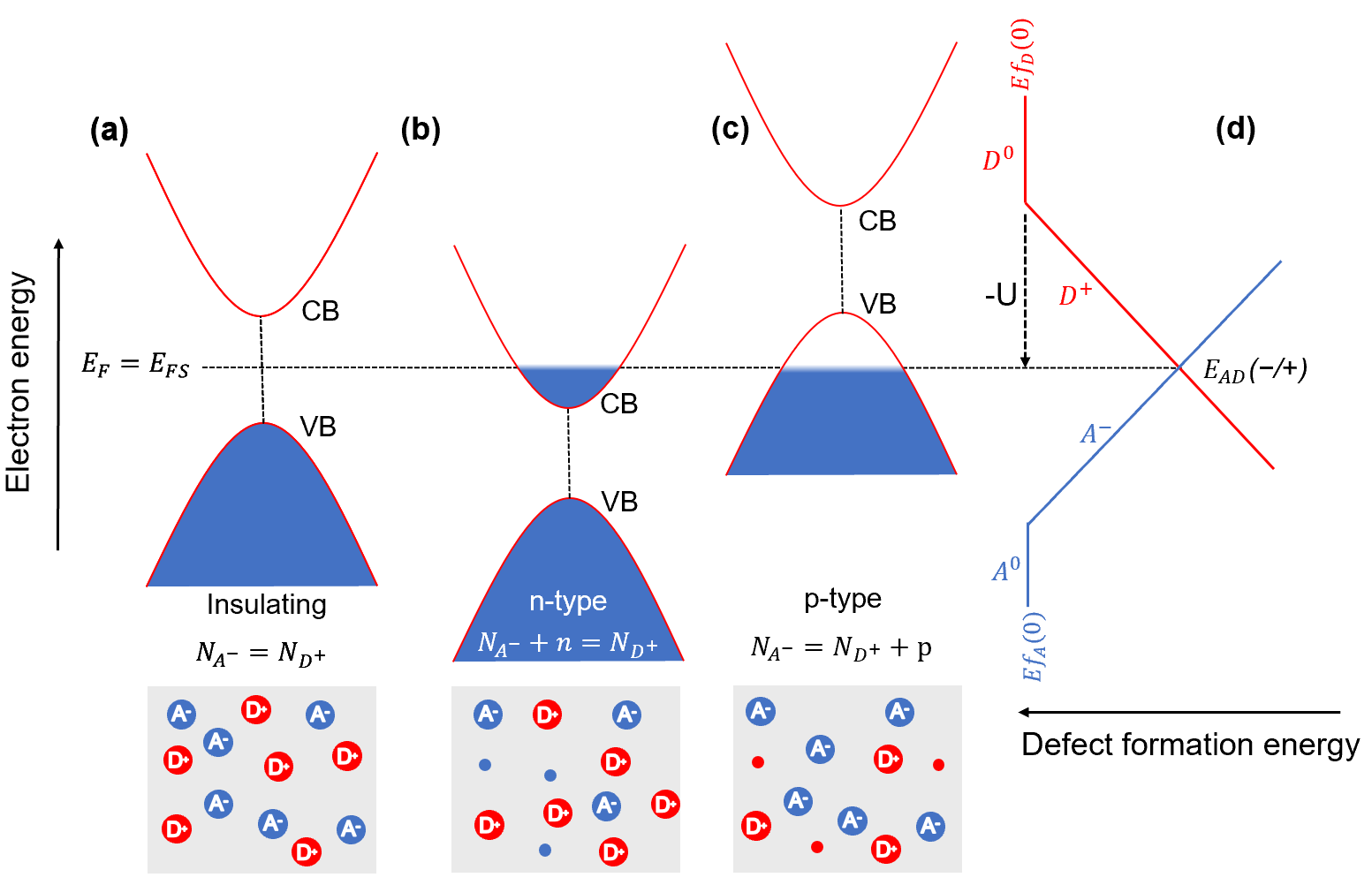}

    \caption{ (a), (b) and (c) Examples of possible three different
    locations of the conduction band (CB) and the valence band (VB) relative
    to $E$\textsubscript{FS} located at 4.9 eV below the vacuum level. (d)
    dependence of the formation energies of amphoteric defects (ADs) in the
   donor and acceptor configuration on the Fermi energy. 
   The effective negative energy $-U$ binds two holes on an amphoteric defect in the donor configuration.}
\label{fig:figure1}    
\end{figure}

\textbf{Figure 1} illustrates three different examples of how the
locations of the extended band edges relative to the $E$\textsubscript{FS}
determine electrical properties of semiconductors in presence of
amphoteric defects (ADs). Thus, as is shown in \textbf{Fig. 1} (a) for
the $E$\textsubscript{FS} located in the band gap the material is
semi-insulating with equal concentration of the amphoteric defects in
the donor and acceptor configuration. This is the case of most standard
semiconductors such as e.g., GaAs and ZnSe or GaN.
However, as is seen in
\textbf{Fig. 1}(b) for high electron affinity semiconductors such as InN
or CdO $E$\textsubscript{FS} pins the Fermi energy in the conduction band
making these materials n-type conducting with majority of ADs in the
donor configuration. Indeed, these two semiconductors are well known to
show large intrinsic n-type conductivities with electron concentrations
reaching mid 10\textsuperscript{20} cm\textsuperscript{-3} range
.\cite{Speaks2010,Jones2006c}  On the opposite side materials with low
ionization energy in which $E$\textsubscript{FS} falls into the valence
band are intrinsically p-type conducting. For example, in the case of
SnTe majority of ADs are in the acceptor configuration and pin
$E$\textsubscript{FS} at about 0.5 eV below the valence band edge
stabilizing the free hole concentration at
\(3 \times 10^{21}\, \text{cm}^{- 3}\).\cite{Nishitani2014}

In superconducting cuprates, the band gap originates from the
metal-insulator transition that splits a partially filled d-band of Cu
into fully occupied charge transfer band (CTB) separated from an empty
upper Hubbard band (UHB) by large charge transfer gap of 1.5 to 2 eV.\cite{Basov2005} 
Despite different origin of the band gap,
incorporation of acceptor-like (donor-like) defects contributes free
holes (electrons) to the CTB (UHB) the same way as in semiconductors. As
will be seen later the situation encountered in superconducting cuprates
corresponds the case depicted in \textbf{Fig. 1} (c) where the
$E$\textsubscript{FS} is located in the charge transfer band (CTB) with
majority of ADs in the acceptor configuration producing conducting free
holes in \ce{CuO2} planes.

\section{Amphoteric Defects in Cuprates}

In general, the electrical properties of non-metallic materials at
elevated temperatures are determined by the intrinsic concentrations of
free carriers thermally excited over a band gap, chemical doping and the
concentration of electrically active lattice defects. Because of the
large energy gap between CTB and UHB,\cite{Basov2005} we assume that
the intrinsic carrier concentrations can be safely ignored at the
annealing or thermal processing temperatures typically used to optimize
superconducting properties of cuprates. Consequently, the conductivity
of cuprates is fully determined by crystal lattice defects and/or by
intentional chemical doping. Chemical doping was used to synthesize
superconducting La\textsubscript{2-x}Ba\textsubscript{x}CuO\textsubscript{4} compound
where divalent Ba atoms partially replacing trivalent La act as
acceptors providing holes to the CTB of the \ce{CuO2} planes.\cite{Bednorz1986} 
On the other hand, in the case of \ce{YBa2CuO_{6 + \delta}} the holes
originate from nonstoichiometry associated with the excess of O, $\delta$. It
is important to note that in both instances an optimization of the
critical temperature required an elaborate processing of the samples at
elevated temperatures with controlled ambient indicating that the
superconducting properties of the samples are determined by a balance
between chemical dopants and intrinsic defects at processing
temperatures.\cite{Chu2015}

Guided by the extensive experience with amphoteric defects in
semiconductors and taking into account that \ce{CuO2} planes
are the only irreplaceable components of all cuprate superconductors it
is reasonable to assume that the amphoteric defects in cuprates are
associated with the ionic Cu-O bonds. These bonds are primarily formed
out of Cu \(d_{x^{2} - y^{2}}\) and O \(p_{x,y}\) orbitals. A removal of
highly electronegative O (metallic Cu) atom from crystal lattice
produces O (Cu) vacancy with the dangling bonds of Cu (O) acting as a
donor (acceptor) with highly spatially localized bound electron (hole).
The total energy of such localized defect is a sum of the electronic
energy and the lattice energy associated with relaxation of the
neighboring atoms. Therefore, the local lattice configuration depends on
the charge state of the defect which, in turn, is determined by the
location of its charge transition state relative to the Fermi energy.

The strong coupling between electronic and lattice degrees of freedom
drives the transformation between donor and acceptor configurations and
defines the amphoteric nature of the dangling bond defects. Thus, as is
schematically illustrated in \textbf{Fig. 2} (b) an O vacancy
(V\textsubscript{O}) is a donor with an electron bound in the dangling
bonds of Cu. The formation energy of the ionized charged donor defect
increases with increasing Fermi energy and is given by
\( Ef_D(E_F) = Ef_D(0) - [E_{D}(0/+) - E_{F} ] \),
where \( Ef_D(0) \) is the formation energy of the
neutral donor and \( E_{D}(0/+)\) is the energy of
the donor charge transition state (\textbf{Fig. 3} (d)). At high enough
Fermi energy the defect can lower its total energy by, shown in
\textbf{Fig. 2}(a), local bond restructuring in which a neighboring Cu
atom moves towards the O vacant site forming
(Cu\textsubscript{O}+V\textsubscript{Cu}) acceptor defect complex with a
hole bound on the O dangling bonds and with the formation energy given
by 
\( Ef_A(E_F) = Ef_A(0) - [E_{F} - E_{A}(-/0) \),
where \(\ Ef_A(E_F) \) is the formation energy of the
neutral acceptor and \( E_{A}(-/0)\) is the energy
of the acceptor charge transition state.

Absent of any other constraints the thermodynamic minimum energy of such
defect system is achieved when the formation energies of the defects in
the acceptor and donor configurations are equal,
\( Ef_{A}( E_{F}) = Ef_{D}( E_{F}) \) i.e. when the Fermi energy is
stabilized at
\(E_{F} = E_{\text{FS}} = E_{\text{AD}}(+/-) = [ Ef_{A}(0) - Ef_{D}(0) - E_{A}(-/0) + E_D(0/+)]/2 \).
In the following, for simplicity we assume
\( Ef_{A}(0) = Ef_{D}(0) \) which
gives
\(E_{F} = E_{\text{FS}} = E_{\text{AD}}(+/-) = [ E_{D}(0/+) - E_{A}(-/0)]/2) \).
In general, it is expected that as is shown in \textbf{Fig. 2}(c) the
bond rearranging transformation between neutral states of the donor and
acceptor configuration requires surmounting an energy barrier,
$E$\textsubscript{b}. The barrier plays an important role as it determines
the temperature and annealing time needed to reach a thermal equilibrium
for ADs and to optimize superconducting properties. Also, it can be a
critical factor affecting the long-term stability of the material at
room temperature.

\begin{figure}
\centering
\includegraphics[width=6in,height=3.18264in]{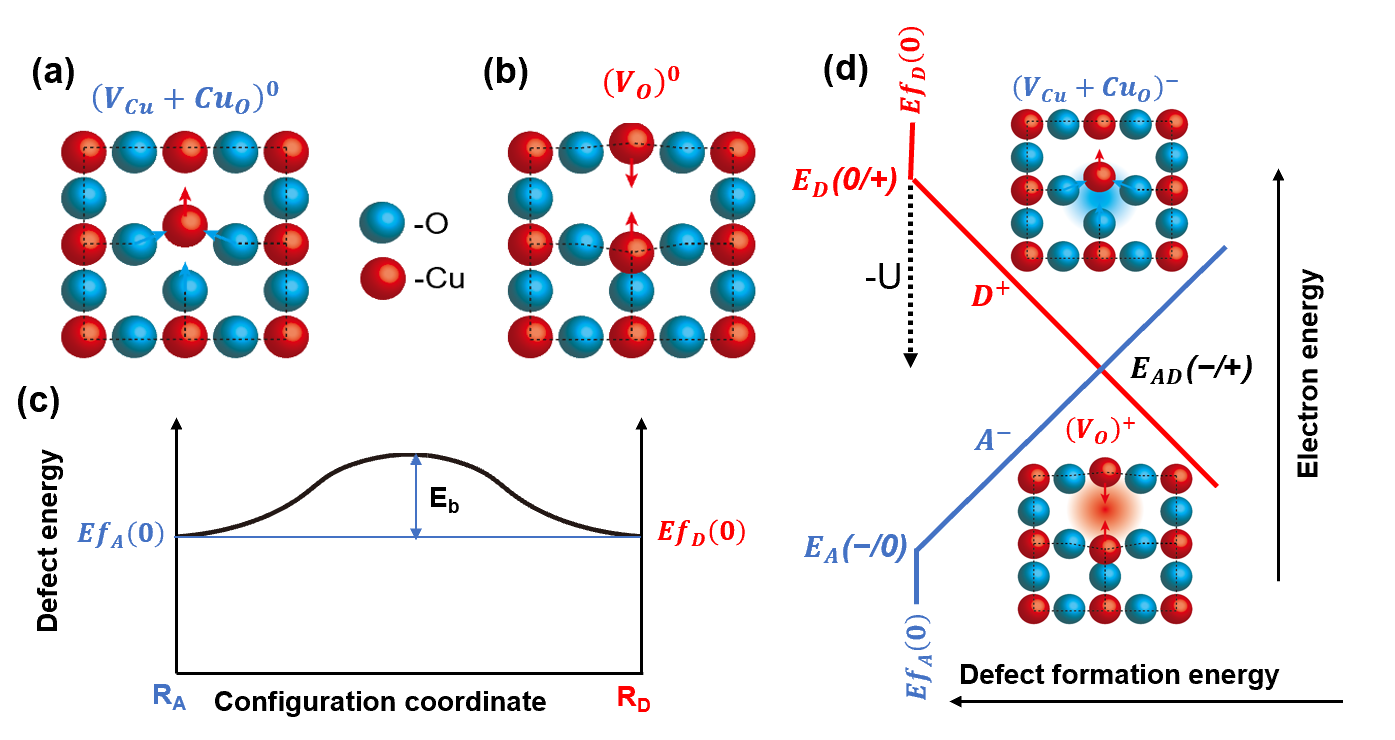}
\caption{Schematic representation of the bond breaking atom
relocation associated with the transformation of an amphoteric defect
between neutral donor (V\textsubscript{O})\textsuperscript{0} (b) and
acceptor (V\textsubscript{Cu}+Cu\textsubscript{O})\textsuperscript{0}
(a) configuration. (c) defect energy dependence on the atom
configuration coordinate with an energy barrier E\textsubscript{b}. (d)
dependence of the amphoteric defects formation energy on the Fermi
energy. The minimum of the total energy is achieved for the Fermi energy
equal to \(E_{\text{AD}}(+/-)\).}
\label{fig:figure2}
\end{figure}


The described above Fermi energy-controlled transformation of the
dangling Cu bonds of a V\textsubscript{O} donor into dangling O bonds of
(Cu\textsubscript{O}+V\textsubscript{Cu}) acceptor applies to an
analogous amphoteric defects system where the dangling O bonds of
V\textsubscript{Cu} acceptor undergo the Fermi energy driven
transformation into dangling Cu bonds of
(O\textsubscript{Cu}+V\textsubscript{O}) donor defect complex. This
again leads to Fermi level driven transformation between donor and
acceptor configuration stabilizing the Fermi energy at the charge
transition state \(E_{\text{AD}}(+/-)\) . It
needs to be emphasized that this thermodynamically driven equilibrium
between free carriers and localized defects is the central concept of
our model of superconductivity in cuprates.

In principle, the stabilization of the Fermi energy can be achieved in a
stoichiometric material by annealing at a high temperature when the
thermal equilibrium concentrations of vacancies on Cu and O sites are
large enough to stabilize the Fermi energy at
\(E_{\text{AD}}(+/-)\). Such high temperature
thermodynamic equilibrium can then be frozen by a rapid cooling to low
temperatures at which the defects cannot undergo the configurational
transformations. Therefore, nonstoichiometry measured by excess or
deficiency of O is not a unique determinant of the carrier concentration
or the concentration of ADs in the donor or acceptor configuration and
cannot be used to evaluate electrical properties of the material.
However, the stabilization of the Fermi level is greatly facilitated in
nonstoichiometric compounds with O or Cu deficiency that has to be accommodated by an increase in the concentration of ADs. This is equivalent to a reduction of the formation energy of neutral ADs, $Ef_A(0)=Ef_D(0)$ (see \textbf{Fig. 2} (d)).

As was discussed in the preceding section the electrical properties of
an ADs containing material are determined by the location of
\(E_{\text{FS}} = E_{\text{AD}}(+/-)\) relative
to the fully occupied valence band or CTB and empty conduction band or
UHB. In superconducting cuprates $E$\textsubscript{FS} is located below
CTB edge leading to the charge balance condition
\(N_{A^{-}\ } = p + N_{D^{+}\ }\) where \(N_{A^{-}\ }\)
(\(N_{D^{+}\ }\)) is the concentrations of negatively (positively)
charged ADs in the acceptor (donor) configuration, p is the
concentration of free holes in the CTB and
\(N_{\text{AD}} = N_{A^{-}\ } + N_{D^{+}\ }\) is the concentration of
the amphoteric defects (ADs). Effects of chemical doping with either
acceptors or donors can be easily incorporated in this picture and will
be discussed in section V of the paper.

\section{Generic nonstoichiometric cuprate superconductor}

The conceptually simplest and the most illuminating cases of high
$T_c$ cuprate superconductors are represented by compounds
such as $\ce{YBa2CuO_{6 + \delta}}$\cite{Wu1987}
 or
Bi\textsubscript{2}Sr\textsubscript{2}CaCu\textsubscript{2}O\textsubscript{8+x}
\cite{Kendziora1996} in which there is no intentional chemical doping
and the concentration of conducting holes and the superconducting
properties are entirely determined by the deviation of the O content
$\delta$, or $x$ from the composition of the parent stoichiometric compound. 
The nonstoichiometric materials are important because they provide the most
straightforward path to explain the main features of our model and to
highlight the role of the amphoteric defects which, as will be argued
later, are solely responsible for the superconductivity in all high
temperature cuprate superconductors.

Here we focus our considerations on a model cuprate superconductor that
is based on the most extensively studied \ce{YBa2CuO_{6 + \delta}} compound
which is a charge transfer insulator for stoichiometric composition \ce{YBa2CuO_{6}} ($\delta = 0$). 
We consider a generic case of single system of amphoteric defects associated with 2-D \ce{CuO2} planes in which free holes originate from ADs acceptors ignoring the role of 1-D CuO chains that are known to affect the charge balance by providing additional holes to the \ce{CuO2} planes. The effects of the
1-D CuO chains will be discussed later in the section V on chemical doping.

\textbf{Figure 3} shows the dependencies of the concentration of free
holes as well as the concentrations of the defects in the $D^{+}$ and
$A^{-}$ configuration on the total concentration of amphoteric
defects, $N_{\text{AD}}$. All the concentrations are measured as
fractions of the total concentration of Cu-O bonds. In these units the
concentrations of amphoteric defects associated with a deviation from
stoichiometry for
\ce{YBa2CuO_{6 + \delta}} is given by
$N_{\text{AD}} = \delta/6$. The corresponding free hole concentrations can be calculated considering that the free holes are localized in 2-D \ce{CuO2} planes with a constant density of
states given by

\begin{equation}\label{Eqn1}
G_{2D} = \frac{4\pi m_{h}^{*}}{h^{3}} = 4.2 \times 10^{14}\left( \frac{m_{h}^{*}}{m_{e}} \right)\text{\ \ }
\left[ \text{eV}^{- 1}\text{cm}^{- 2} \right]\,
\end{equation}
\noindent
where \(m_{h}^{*}\) and \(m_{e}\) are the hole effective mass in the CTB
and the free electron mass, respectively. Adopting the plane unit cell
dimensions for YBCO $a$ = 0.38 nm and $b$ = 0.39 nm yields the density of states as a fraction of the concentration of CuO bonds:\cite{Beno1987}

\begin{equation}\label{Eqn2}
G_{2Dfr} = 0.63\left( \frac{m_{h}^{*}}{m_{e}} \right)\text{\ \ }
\left\lbrack \text{eV}^{- 1} \right\rbrack.
\end{equation}
\noindent
Since the interplanar lattice constant $c$ = 1.17 nm then the equivalent 3-dimensional density of states is given by

\begin{equation}\label{Eqn3}
G_{3D} = \frac{G_{2D}}{c} = 3.5 \times 10^{21}\left( \frac{m_{h}^{*}}{m_{e}} 
\right)\text{\ \ }\left[ \text{eV}^{- 1}\text{cm}^{- 3} \right].
\end{equation}

As is illustrated in \textbf{Fig. 3} for the charge transition state
located in the CTB an initial increase of the non-stoichiometry
\(N_{\text{AD}}\)produces amphoteric defects in the acceptor
configuration that contribute free holes to the CTB i.e
\(p = N_{A^{-}\ }\). The free holes frustrate the antiferromagnetic
order of Cu magnetic moments leading to a decrease of the
antiferromagnetic coupling and a reduction in the Neel temperature.
Furthermore, the increase of the free hole concentration downward shifts
the Fermi energy in the CTB until it reaches the charge transition state
\(E_{\text{AD}}\left( - / + \right)\). Since this Fermi energy
corresponds to the equal formation energy of the amphoteric defects in
the acceptor and donor configuration an additional increase of
\(N_{\text{AD}}\) leads to formation of equal concentrations of
\(A^{-}\) and \(D^{+}\) that compensate each other stabilizing the Fermi
energy at
\({E_{F} = E}_{\text{FS}} = E_{\text{AD}}\left( - / + \right)\) and the
corresponding hole concentration at \({p = p}_{\text{s\ }}\). This is a
critical transition point as a further increase of \(N_{\text{AD}}\)
increases the concentration of the defects in the \(D^{+}\)
configuration which, as will be shown later are responsible for the
superconductivity. It is easy to see that the hole concentration, $p$ and
the concentrations of ADs in the acceptor, \(N_{A^{-}\ }\)and donor,
\(N_{D^{+}\ }\) configuration are given by:

\begin{subequations}
\begin{align}                    
p &= N_{\textrm{AD}}\quad \textrm{for }\, N_{\text{AD}} \leq p_s  \\
p &= p_{s}\quad \textrm{for } \, N_{\text{AD}} > p_s \nonumber \\
N_{A^{-}} &= N_{\text{AD}}\quad \textrm{for }\, N_{\text{AD}} \leq p_{s} \\
N_{A^{-}} &= \frac{( N_{\text{AD}} + p_{s})}{2}\quad \textrm{for } N_{\text{AD}} > p_{s} \nonumber\\
\intertext{and}
N_{D^{+}} &= 0\quad\textrm{for }\, N_{\text{AD}} \leq p_{s}\\
N_{D^{+}} &= \frac{( N_{\text{AD}} - p_{s})}{2}\quad \textrm{for }\,N_{\text{AD}} > p_{s} \nonumber
\end{align}
\end{subequations}

The equilibrium condition between amphoteric defects in the acceptor,
\(A^{-}\) and donor, \(D^{+}\) configurations for
\(p_{\text{s\ }} \leq N_{\text{AD}}\) is achieved at high temperature
thru a capture of two charges. Capture of two holes that transforms
\(A^{-}\) into \(D^{+}\) is described by the reactions,

\begin{subequations}
\begin{align}
A^{-} + h &\longrightarrow A^{0} \\
A^{0} &\Rightarrow D^{0} \\
D^{0} + h &\longrightarrow D^{+}
\end{align}
\end{subequations}

\noindent
with an analogous set of reactions for transformation of a \(D^{+}\)
into \(A^{-}\) by sequential capture of two electrons:

\begin{subequations}
\begin{align}
D^{+} + e &\longrightarrow D^{0}\\
D^{0} &\Rightarrow A^{0}\\
A^{0} + e &\longrightarrow A^{-}
\end{align}
\end{subequations}

\noindent
Although the capture of the first hole (electron) in the reaction 5(a)
(6(a)) requires the excitation energy
\(U = E_{\text{AD}}\left( - / + \right) - E_{A}\left( - /0 \right)\) (
\(U = {E_{D}\left( 0/ + \right) - E}_{\text{AD}}\left( - / + \right)\))
this energy is regained by adding another hole (electron) to the defect
in the reaction 5(c) (6(c)). This is a very important property of the
ADs as it drives the reaction,
\({2AD}^{0} \longrightarrow {D^{+} + A}^{-}\), in which two neutral ADs
lower their energy by exchanging charges and producing \(D^{+}\) defect
with localized pair of equivalent holes and an \(A^{-}\) defect with
localized pair of equivalent electrons bound by the energy \(- U\).
These properties of ADs create a unique physical system of free holes
whose Fermi energy is resonant with the energy level of hole pairs
localized on the randomly distributed, isolated \(D^{+}\) donor defects.

In general terms a similar system was previously envisioned as a
possible scheme to explain superconductivity in cuprates
 where it was argued that the very small coherence
length indicated that the superconductivity in these materials has to be
mediated by a field, \(\Phi\) of highly localized bosons.\cite{Friedberg1989a}
Here we show
that this condition is satisfied by an ensemble of randomly distributed
\(D^{+}\) defect centers localizing two holes with the binding energy
$U$ and acting as a boson field $\Phi$ leading to the transition,

\begin{equation}
 2h \longrightarrow \Phi \longrightarrow 2h,   
\end{equation}

\noindent
in which localized zero momentum boson field of $D^{+}$ defect centers
forces two mobile holes with equal and opposite momenta to form Cooper
pairs that at low enough temperature evolve into a coherent
superconducting phase. It should be noted that because of the planar
nature of the Cu-O bonds the localization of the \(D^{+}\) defects is
anisotropic with the spatial extent restricted to few nearest neighbors
in the \ce{CuO2} plane and with even stricter confinement in
the direction perpendicular to the \ce{CuO2} plane. This is
fully consistent with a small coherence length of
$\xi \approx 2.5$ nm in the \ce{CuO2} plane and an
extraordinarily small $\xi \approx 0.8$ nm perpendicular to the plane
found in cuprate superconductors.\cite{Chaudhari1987} Also, as was
argued in ref. \onlinecite{Friedberg1989a} the coupling between the fermion and
boson systems does not depend on the internal structure of the localized
centers that form the boson field. This means it is not affected by the
microscopic structure of the $D^{+}$ defects localizing hole pairs.

It is obvious from the above considerations that cuprates cannot be
considered classical metals with simple metallic bonds and a free
carrier gas described by the Fermi liquid theory. Instead, the electrical
properties of cuprates are determined by interactions between free hole
gas and the Cu-O bonding structure that is critically affected by the
material stoichiometry. Thus, in this case the Cu and/or O dangling
bonds play a role of so called ``internal coordinates'' which, as has
been hypothesized before, could play a key role in determining normal and
superconducting properties of unconventional superconductors.\cite{Ting1980}

An explicit theory of superconductivity in a system with localized
negative-$U$ centers was previously developed to consider the
experimentally observed enhancement of the critical temperature in
superconducting eutectic Al-Si alloys.\cite{Simanek1979,Ting1980a} It was
shown that interaction of the Al metal electrons with negative-$U$ centers
of amorphous phase of Si leads to enhancement of the critical
temperature of Al metal.\cite{Ting1980a} The theory has also
yielded an expression for the critical temperature of a free fermion gas
interacting with localized negative-$U$ bosons. Adopting this approach to
our presently considered system allows expressing the critical
temperature for superconducting transition in terms of equations where
key parameters can be derived from the AD model.

The critical temperature $T_c$ for superconducting
transition is obtained solving the equations,\cite{Ting1980a}

\begin{equation}\label{Eqn8}
    1 = \left| U_{\text{eff}} \right|\Gamma\left\lbrack \varphi\left( T_{c},E_{r}^{2} \right) 
    - \varphi\left( T_{c},E_{r}^{2} + \left| v_{\text{AD}}^{2} \right|N_{D +} \right) \right\rbrack
\end{equation}
\noindent
with

\begin{equation}
\varphi\left( T_{c},E_{r}^{2} \right) = 
\frac{1}{\pi\left(\Gamma^{2} + E_{r}^{2} \right)} 
\left[\ln\frac{2\gamma\left( \Gamma^{2} + E_{r}^{2} \right)^{\frac{1}{2}}}{\pi T_{c}} - \frac{\Gamma}{T_{c}}\tan^{- 1}\left(\frac{E_{r}}{\Gamma} \right) \right],
\end{equation}

\begin{equation}
U_{\text{eff}} = 
\frac{U}
{1 + \left( \frac{U}{\pi E_{r}} \right)\tan^{-1}\left( \frac{E_{r}}{\Gamma} \right)} < 0,
\end{equation}

\noindent
and $\Gamma = \pi\left| v_{\text{AD}}^{2} \right|G\left( 0 \right)$, 
where \(N_{D^{+}\ }\) is the concentration of localized hole pairs equal
to the concentration of the ADs in the donor configuration,
\(E_{r} = E_{\text{AD}}(+/-) - E_{\textrm{F}}\) is the
energy of the charge transition state of ADs between donor and acceptor
configuration relative to the Fermi energy, $U$ is the negative energy
localizing two holes on the \(D^{+}\) defects,
\(\left| v_{\text{AD}}^{2} \right|\) is the parameter describing the
strength of the coupling between free holes and hole pairs localized on
the \(D^{+}\) defects, \(G\left( 0 \right) = G_{2Dfr}\) is the density
of states of free holes at the Fermi energy and \(\gamma\) is 
Euler's constant.

The dependencies of the free hole concentration (\(p\)) and the
concentrations of ADs in the acceptor (\(N_{A^{-}\ }\)) and donor
configuration (\(N_{D^{+}\ }\)) on the concentration of the ADs
(\(N_{\text{AD}}\)) given by relations 4(a) to 4(c) are shown in
\textbf{Fig. 3} assuming \(p_{\text{s}} = 0.05\). Using Eqs. (1) and
(2) with the hole effective mass \(m_{h}^{*} = 0.62m_{e}\) this
\(p_{\text{s\ }}\) corresponds to 2-D hole concentration of
$3.4 \times 10^{13}\, \text{cm}^{- 2}$ and the Fermi energy located at
0.13 eV below the CTB edge. The solid black line represents critical
superconducting temperature $T_c$ obtained by solving Eqs.
(8) to (10). As expected, $T_c$ is increasing with the
concentration of the localized hole pairs, $N_{D^{+}}$. All the
calculations in this paper were performed with a single set of
parameters, \(U = - \ 0.52\) eV,
\(G\left( 0 \right) = 0.39 = 2.6 \times 10^{14}\ \text{eV}^{- 1}\text{cm}^{- 2}\),
\(\left| v_{\text{AD}}^{2} \right| = 0.16\) eV and \(\Gamma = 0.2\) eV.
Considering the number of the parameters these is not a unique set. The
parameter space was investigated and the final set of parameters was
chosen to yield a reasonable overall agreement of the calculated
quantities with typical experimental data.

The $T_c$ plotted in \textbf{Fig. 3}(h) was calculated for
\(E_{r} = 0\), i.e for the Fermi energy resonant with the charge
transition state, \(E_{\text{AD}}(+/-))\) between
donor and acceptor configuration. To illustrate the resonant character
of the superconducting coupling, the lowest panel (Fig. 3 (i-k)) shows
the dependence of $T_c$ on the deviation
\( E_r/\Gamma \) of the Fermi energy from the resonance,
calculated for three different concentrations of ADs. It is seen that a
departure from the resonance condition leads to a rapid reduction of
$T_c$. These results highlight the importance of the high
temperature processing whose objective is not only to generate free
holes but more importantly to create a thermodynamic equilibrium with
maximized concentration of isolated boson \(D^{+}\) centers and the
Fermi energy as close to resonance with
\(E_{\text{AD}}(+/-)\) as possible. It also raises
an interesting issue of the extent to which the resonant conditions
obtained at high processing temperature are still preserved at low
temperatures at which superconductivity is measured.

\begin{figure}
  \centering
  \includegraphics[width=5.93941in,height=4.46154in]{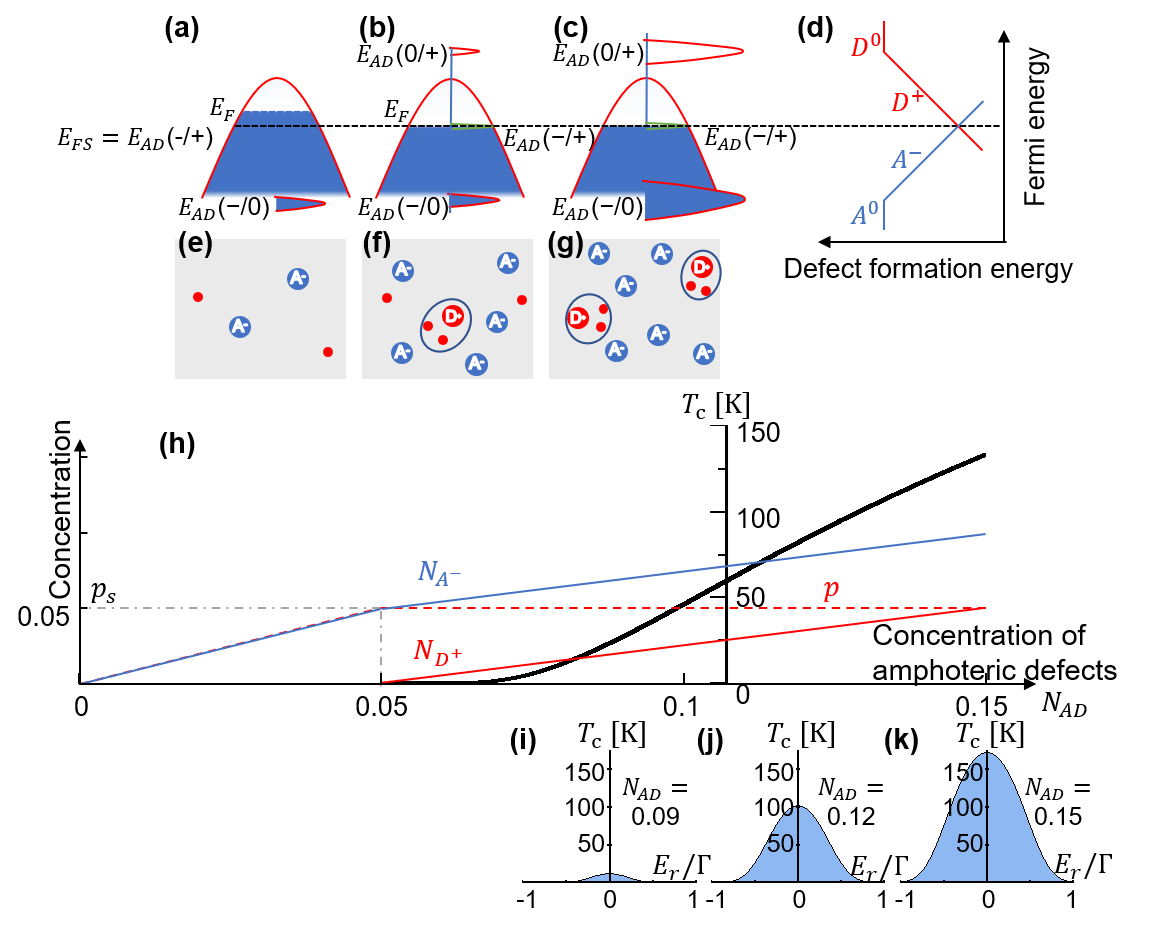}
  \caption{(a), (b) and (c) downward shift and stabilization of
the Fermi energy in the CTB with increasing concentration of ADs with
the formation energy shown in (d); schematic representation of the
relationship between free holes and ADs in the acceptor and donor
configuration (e), (f) and (g). Two holes localized on the donor defect
force formation of a pair of free holes with opposite momenta (f) and
(g). Dependence of the free hole concentration along with concentrations
of the defects in the acceptor and donor configurations as functions of
the total concentration of amphoteric defects (e). The black solid line
shows the calculated critical temperature $T_c$. The bottom
panel (i), (j) and (k) shows the dependence of the $T_c$ on
the deviation from the resonance condition
${E_{F} = E}_{\text{AD}}(+/-)$ for three
different concentrations $N_{\text{AD}}$.}
\label{fig:figure3}
\end{figure}

The calculations in \textbf{Fig. 3} showing the dependence of
$T_c$ on \(N_{D^{+}}\) were carried out assuming an
idealized system of randomly distributed, noninteracting defects that
are in a stable thermodynamic equilibrium that has been achieved during
high temperature processing. However, since maintaining the
superconducting resonance coupling condition requires that
\( N_{A^{-}} = p_{\text{s}} + N_{D^{+}}\) this means that
attaining a higher $T_c$ through an increase in
\(N_{D^{+}}\) has to be compensated by equal increase in the
concentration of the negatively charged ADs in the acceptor
configuration, \(N_{A^{-}}\) . This leads to a decrease of the average
distance between \(A^{-}\) and \(D^{+}\) centers and an increased
probability for a passivation by formation of a neutral \({A^{-}D}^{+}\)
complexes that deactivate \(D^{+}\) resulting in lower
$T_c$.

To account for this effect, we adopt a simplified approach based on the
concept of metal-insulator transition in doped semiconductors in which
we assume that there is a critical concentration of ADs in the acceptor
configuration, \(N_{A^{-}}^{\text{cr}}\) at which all \(D^{+}\) will be
passivated resulting in an abrupt cut-off of the superconductivity. To
evaluate the effects of the passivation on the macroscopic
superconductivity we consider how spatial fluctuations of the defect
concentration, \(N_{\text{AD}}\) affect \(N_{D^{+}\ }\) and thus also
the critical temperature. A spatially inhomogeneously broadened
concentration of the \(N_{D^{+}\ }\) is given by

\begin{equation}
N_{D^{+}}^{\text{br}} = \frac{1}{\sigma\sqrt{2\pi}}\int_{- \infty}^{\infty}{N_{D^{+}}^{'}\exp\lbrack - \frac{\left( N_{D^{+}}^{'} - N_{D^{+}} \right)^{2}}{2\sigma^{2}}}\rbrack dN_{D^{+}}^{'}
\end{equation}
\noindent
where \(\sigma\) is the broadening parameter,

\begin{equation*}
N_{D^{+}} = \frac{( N_{\text{AD}} - p_{s})}{2}\quad \textrm{for }\,N_{\text{AD}} < 2N_{A^{-}}^{\text{cr}} - p_{s}
\end{equation*}

\noindent
and

\begin{equation*}
N_{D^{+}} = 0\quad \textrm{for }\, N_{\text{AD}} \geq 2N_{A^{-}}^{\text{cr}} - p_{s}
\end{equation*}

\begin{figure}
  \centering
   \includegraphics[width=5.9546in,height=3.15957in]{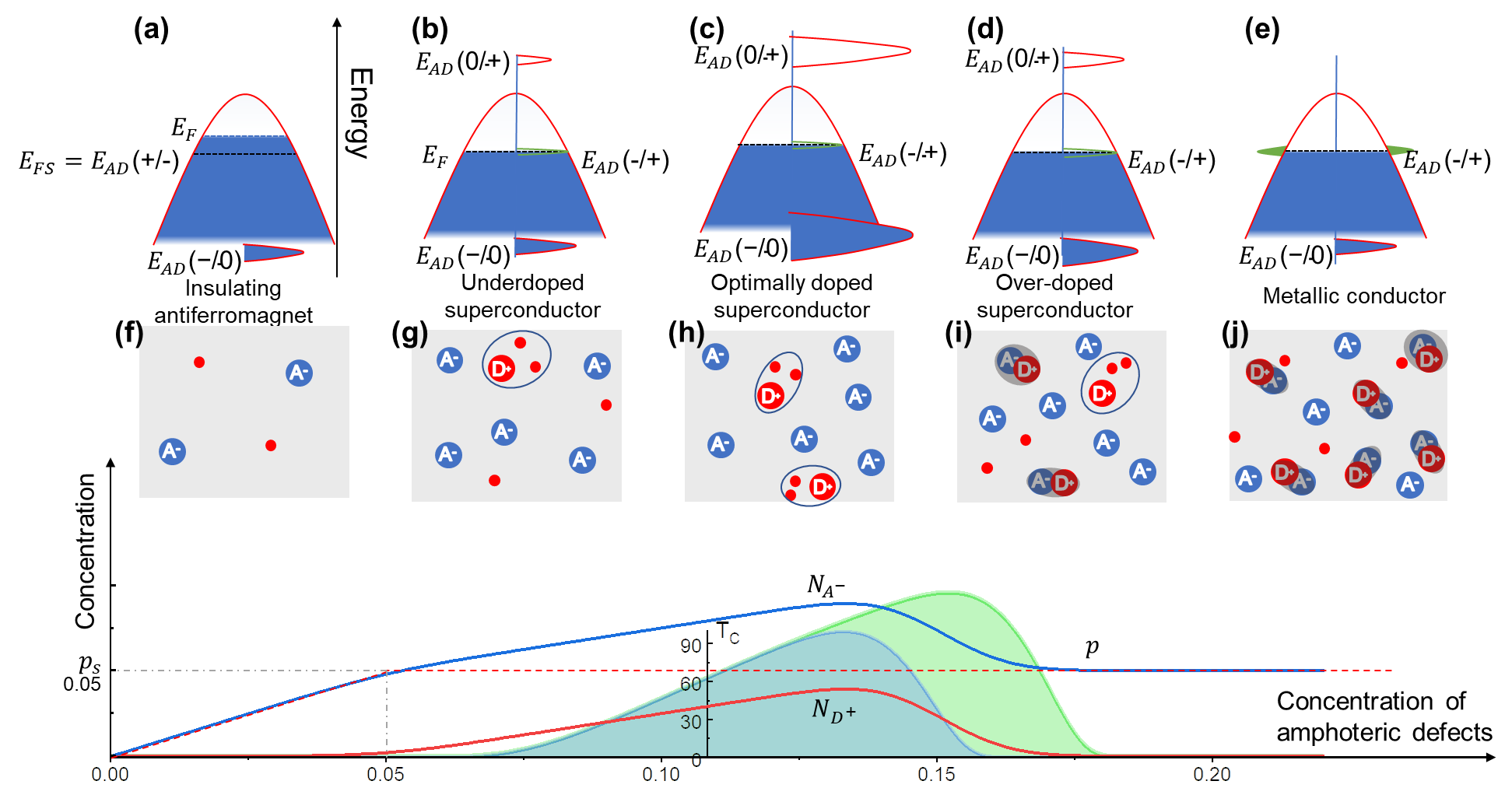}
   \caption {(bottom panel) Dependence of critical temperature
$T_c$ on the concentration of ADs for two different cut-off
densities \(N_{A^{-}}^{\text{cr}}\) = 0.1 (dark green) and 0.11 (light
green). The plots of \(N_{A^{-}}\) and \(N_{D^{+}}\) represent the
former case. Middle panel illustrates schematically the evolution of the
material properties from high resistivity antiferromagnet with all ADs
in the acceptor configuration (f) to underdoped superconductor with
increasing concentration of the localized boson pairs (g), to optimally
doped material with maximum density of boson pairs (h), to overdoped
superconductor with a fraction of \(D^{+}\) passivated by \(A^{-}\)
defects (i) finally to non-superconducting metallic conductor with all
\(D^{+}\) passivated by \(A^{-}\) (j). The top panel shows the
corresponding dependence of the Fermi energy in the CTB and associated
change in the density of states of defects in different configurations.}
\label{fig:figure4}
\end{figure}

The spatially broadened concentration of the AD donors
\(N_{D^{+}}^{\text{br}}\) given by Eq. (11) can then be used to
calculate the critical temperature. \textbf{Figure 4} shows the
dependence of superconducting transition temperatures $T_c$
on the \(N_{\text{AD}}\) calculated using two different critical
concentrations, \(N_{A^{-}1}^{\text{cr}} = 2p_{s} = 0.1\) and
\(N_{A^{-}2}^{\text{cr}} = 2.2p_{s} = 0.11\ \) corresponding to
\(N_{AD1}^{\text{cr}} = 0.15\ \)and \(N_{AD2}^{\text{cr}} = 0.17\). It
is seen that incorporation of the broadening for the defect distribution
leads to a gradual decrease of the \(N_{D^{+}}\) and thus also
$T_c$ at large \(N_{\text{AD}}\). This results in a
characteristic dome-like shape in the \(N_{\text{AD}}\) dependence of
$T_c$ with five doping regions. In the first region,
\(N_{\text{AD}} < 0.05\ (\delta < 0.3)\) the material has low hole
concentration, \({p = N}_{\text{AD}}\) which is consistent with the
high resistivity antiferromagnetic region experimentally observed in all
high $T_c$ cuprates. For \(N_{\text{AD}} \geq p_{s}\) the
hole concentration stabilizes at p\textsubscript{s}. The stabilization
is accompanied by formation of the hole boson pair \(D^{+}\) centers and
the onset of the superconductivity. In this concentration range commonly
termed as the underdoped region the $T_c$ is monotonically
increasing with \(N_{\text{AD}}\). At even higher \(N_{\text{AD}}\)
the $T_c$ begins to decrease when the hole boson pair
\(D^{+}\) centers become progressively passivated by \(A^{-}\) defects.
This range of concentration is referred to as an overdoped region.
Finally, $T_c$ goes to zero when all \(D^{+}\) centers are
passivated and, as is shown in Fig. 4 (e) the defects form a narrow band
of extended states at the charge transition state
\(E_{\text{AD}}(+/-)\) leading to an increased
density of states at the Fermi energy and the onset of metallic
conductivity. It is important to note that in this simple picture the
optimum doping region just corresponds to the maximum $T_c$
in the transition from the underdoped region with increasing \(D^{+}\)
and $T_c$ to the overdoped region with the onset of the
defect passivation, decreasing concentration of isolated \(D^{+}\) and
thus also $T_c$. Actually, the dependence of \(N_{D^{+}}\)
and $T_c$ on \(N_{\text{AD}}\) in the optimum doping
region is more complicated since, as is shown in the bottom panel of
\textbf{Fig. 3} $T_c$ is strongly affected by any deviation
from \(E_{\text{r\ }} = 0\) resonance that in turn depends on the sample
preparation conditions. This will be further discussed in section VI of
the paper.

Besides the good description of the dome like phase diagram the AD model
accounts for some other well experimentally established but poorly
understood features of superconducting cuprates. One of the mysterious
characteristics of cuprates is the universal dependence of the critical
temperature on the superconducting phase or superfluid density measured
by muon-spin-relaxation\cite{Uemura1989}  where the same simple
relationship between $T_c$ and superfluid density was found
for different cuprates in the underdoped\cite{Uemura1989} and to
some extent in the overdoped \cite{Niedermayer1993} region. This
relationship can be easily understood considering that, as is shown
\textbf{Fig. 3}, the superfluid density and the critical temperature are
proportional to \(N_{D^{+}}\) which is the only variable parameter in
the Eqs. (8) to (10) to calculate $T_c$. All the other
parameters are determined by properties of 2-D \ce{CuO2}
planes that are very similar for all cuprates leading to shown in
\textbf{Fig. 4} universal $T_c$ dependence on the doping
for all cuprates in the underdoped region. It has been shown that the
difference in the maximum $T_c$ in the optimally doped
region can be attributed to the number of \ce{CuO2} planes
per material formula in different cuprates. Thus, it has been
demonstrated that in 
\ce{HgBa2Ca_{m-1}Cu_mO_y} the maximum $T_c$ is increasing from $T_c$ = 94
K for \(m = 1\) to 134 K for \(m = 3\).\cite{Schilling1993,Putilin1993} 
This again has a simple explanation as with increasing m the same
concentration of ADs corresponds to larger average distance between
defects and thus also larger $N_{A^{-}}^{\text{cr}}$ which as is shown
in \textbf{Fig. 3} shifts the optimum doping to larger
$N_{\textrm{AD}}$ and higher $T_c$. The decrease in
$T_c$ for $m > 3$ could be associated with reduced hole confinement in the \ce{CuO2} planes.

\section{Chemical doping}

The superconductivity in cuprates was first discovered in intentionally
chemically doped
La\textsubscript{2-x}Ba\textsubscript{x}CuO\textsubscript{4}
\cite{Bednorz1986} where Ba atoms partially substituting La atoms
contribute free holes to the CTB. The large majority of subsequent work
was done on p-type materials although in most instances the p-type
conductivity was realized by varying the O content rather than chemical
doping. The dominant role of nonstoichiometry was clearly demonstrated
in the case of single crystals of
Bi\textsubscript{2}Sr\textsubscript{2}CaCu\textsubscript{2}O\textsubscript{8+x}
where \cite{Kendziora1996} the properties of the crystals were varied
across the entire superconducting dome by increasing O content, x. Also,
it has turned out that more elaborate annealing conditions in O-rich
environment can also produce superconductivity in
La\textsubscript{2}CuO\textsubscript{4}, which is the parent compound
for the extensively studied
La\textsubscript{2-x}Ba\textsubscript{1-x}CuO\textsubscript{4} and
La\textsubscript{2-x}Sr\textsubscript{1-x}CuO\textsubscript{4}
chemically doped with Ba or Sr.\cite{Grant1987} These findings
indicated that the chemical p-type doping simplifies sample preparation
conditions to achieve optimum critical temperature but is not required
for superconductivity.

A critical advancement in the research on superconducting cuprates was
an unanticipated discovery of superconductivity in materials chemically
doped with donors.\cite{Tokura1989} It was found that, as expected,
Ln\textsubscript{2-x}Ce\textsubscript{x}CuO\textsubscript{4} in which
four valent Ce partially replaces trivalent lanthanide atoms (Ln=Pr, Nd)
is a non-superconducting electron conductor. Surprisingly and somewhat
accidentally it was found that the compounds become superconducting when
subjected to high temperature annealing under O-deficient conditions.\cite{Tokura1989} Interestingly the properties of these n-type
superconducting cuprates exhibit certain qualitative similarities with
the p-type counterparts. Most prominently they show a dome like
dependence of the critical temperature on the doping although, with much
lower critical temperature and a narrower superconducting doping range.\cite{Armitage2010,Fournier2015} 
A striking difference between these two types of superconductors is that producing superconductivity in n-type
cuprates always requires more elaborate and often extreme annealing
procedures\cite{Brinkmann1997}  with very specific, well controlled O
ambient whereas in p-type cuprates a higher $T_c$
superconductivity can be realized with much simpler sample preparation
procedures. It should be emphasized however that in both instances
optimization of superconducting properties involves sample processing
aimed at varying the O content.\cite{Krockenberger2013,Brinkmann1995}

An important and mysterious feature of the n-type doped cuprates is that
they do not consistently show an n-type normal state
conductivity.\cite{Wang1991} In addition, there are experimental
indications that holes rather than electrons are responsible for
superconductivity in these materials.\cite{Wang1991,Matsumoto2009,Dagan2007} Comparing the main features of the two types of chemically doped superconducting
cuprates it can be concluded that in contrast to p-type n-type cuprates
are much more difficult to synthesize, have inferior superconducting
properties and do not exhibit specific, well-defined type of normal
state conductivity.

Considering the discussed in section II bipolar nature of the amphoteric
defects in semiconductors it would be tempting and most straightforward
to assume that, as is shown in \textbf{Fig. 1} (a), chemical n-type
doping stabilizes the Fermi level at the charge transition state located
in the conduction band or UHB and the superconducting phase of free
electrons is associated with formation of electron Cooper pairs coupled
to electron pairs localized on \(A^{-}\) centers. However, this would
require a large, comparable to the band gap, downward shifts of UHB edge
energy relative to the charge transition state energy,
\(E_{\text{AD}}(+/-)\). It is an unlikely
scenario, as all these energies are defined by the nature of the highly
localized Cu-O bonds in \ce{CuO2} planes that are not
significantly different for n- and p-type doped cuprates. Also, direct
x-ray photoelectron spectroscopy measurements found only small
differences in the locations of the Fermi energy in p-type compared with
n-type superconducting cuprates.\cite{Allen1990} Therefore, there is
a need to envision the case of a cuprate doped with chemical donors
contributing electrons to the UHB but with the charge transition state
still located in the CTB.

In order to evaluate how the chemical doping affects the
superconductivity we consider a prototypical generic cuprate chemically
doped with either acceptors or donors. Again, the properties of the
material are determined by interactions between the intentionally
introduced dopants and the ADs whose concentration \(N_{\text{AD}}\)
depends on the material stoichiometry. As is shown in the upper panel of
Fig. 5 chemical doping of a stoichiometric cuprate with \(N_{a}\)
acceptors (\(N_{d}\) donors) produces p-type (n-type) conductor with
\({p = N}_{a}\) holes (\({n = N}_{d}\) electrons) placing the Fermi
energy in the CTB (UHB). In both cases increasing the nonstoichiometry
and thus also \(N_{\text{AD}}\) results in a downward shift of the Fermi
energy towards \(E_{\text{AD}}(+/-)\). Once this
level is reached the charge balance between chemical dopants and ADs in
the donor and acceptor configuration is given by:

\begin{subequations}
\begin{align}
N_{A^{-}} &= \frac{N_{\text{AD}} +  (p_{s} - N_{a^-})}{2}  \\
N_{A^{-}} &= \frac{N_{\text{AD}} -  (p_{s} - N_{a^-})}{2}
\end{align}
\end{subequations}
\noindent
for doping with \(N_{a}\) chemical acceptors, and

\begin{subequations}
\begin{align}
N_{A^{-}} &= \frac{N_{\text{AD}} + (p_{s} + N_{d^+})}{2}\\
N_{D^{+}} &= \frac{N_{\text{AD}} - (p_{s} + N_{d^+})}{2}
\end{align}
\end{subequations}

\noindent
for doping with $N_{d}$ of chemical donors. Here we assume that that
the charge transition state for chemical acceptors (donors) lies well
below (above) the CTB edge and all of the dopants are ionized.

The concentrations of ADs in the acceptor and donor configurations as
functions of \(N_{\text{AD}}\) given by Eqs. (12) and (13) are shown in
\textbf{Fig. 5} with the right-hand side representing p-type and the
left-hand side n-type doping. The same chemical doping concentration,
\(N_{a} = N_{d} = 0.4p_{s}\) was assumed for both cases. It should be
noted however that the main conclusions of the following considerations
are not affected by the chemical doping levels. The chief effect of the
chemical doping is to change the balance between concentration of ADs in
the acceptor (\(A^{-}\)) and donor (\(D^{+}\)) configuration with p-type
(n-type) doping reducing (increasing) \(N_{A^{-}}\) and increasing
(reducing) \(N_{D^{+}}\). This has a direct effect on the
superconductivity because $T_c$ is proportional to
\(N_{D^{+}}\) and as seen in the \textbf{Fig. 5} p-type (n-type)
chemical doping reduces (increases) the concentration of
\(N_{\text{AD}}\) (nonstoichiometry) required for the onset of the
superconducting dome shifting it from \(N_{\text{AD}} \approx 0.05\) for
p-type to \(N_{\text{AD}} \approx 0.09\) for n-type doping. Furthermore,
since the p-type (n-type) chemical doping decreases (increases) the
concentration of \(N_{A^{-}}\) it also has an effect on the termination
of the $T_c$ dome at high \(N_{\text{AD}}\) where, as has
been shown in \textbf{Fig. 4} the decrease of $T_c$ is
associated with passivation of \(D^{+}\) by \(A^{-}\) centers.
\textbf{Figure 5} shows that adopting the same value
\(N_{A^{-}}^{\text{cr}} = {2p}_{s}\) for both p-type end n-type doping
terminates $T_c$ at \(N_{\text{AD}} \approx 0.18\) for
p-type and at \(N_{\text{AD}} \approx 0.13\) for n-type doping. This
distinct asymmetry in the maximum $T_c$ and the shape of
the phase diagram for p-type and n-type doping is in a remarkable
qualitative agreement with existing experimental
data.\cite{Armitage2010} The lowest panel in \textbf{Fig. 5} shows
color-coded two dimensional maps of the dependence of $T_c$
on the \(N_{\text{AD}}\) and the deviation of the Fermi energy from the
resonance condition
\( E_{r}/\Gamma = [E_{F} - E_{\text{AD}}(+/-)]/\Gamma = 0 \).
It is seen that close to the optimum doping $T_c$ becomes
less sensitive to the deviation of the Fermi energy from the resonance.
Therefore, as will be discussed later, it is possible that with specific
material processing conditions the optimum $T_c$ can be
achieved under off resonance conditions (\(E_{r} \neq 0\)) but with
higher \(N_{D^{+}}\).

\begin{figure}
  \centering
\includegraphics[width=5.96709in]{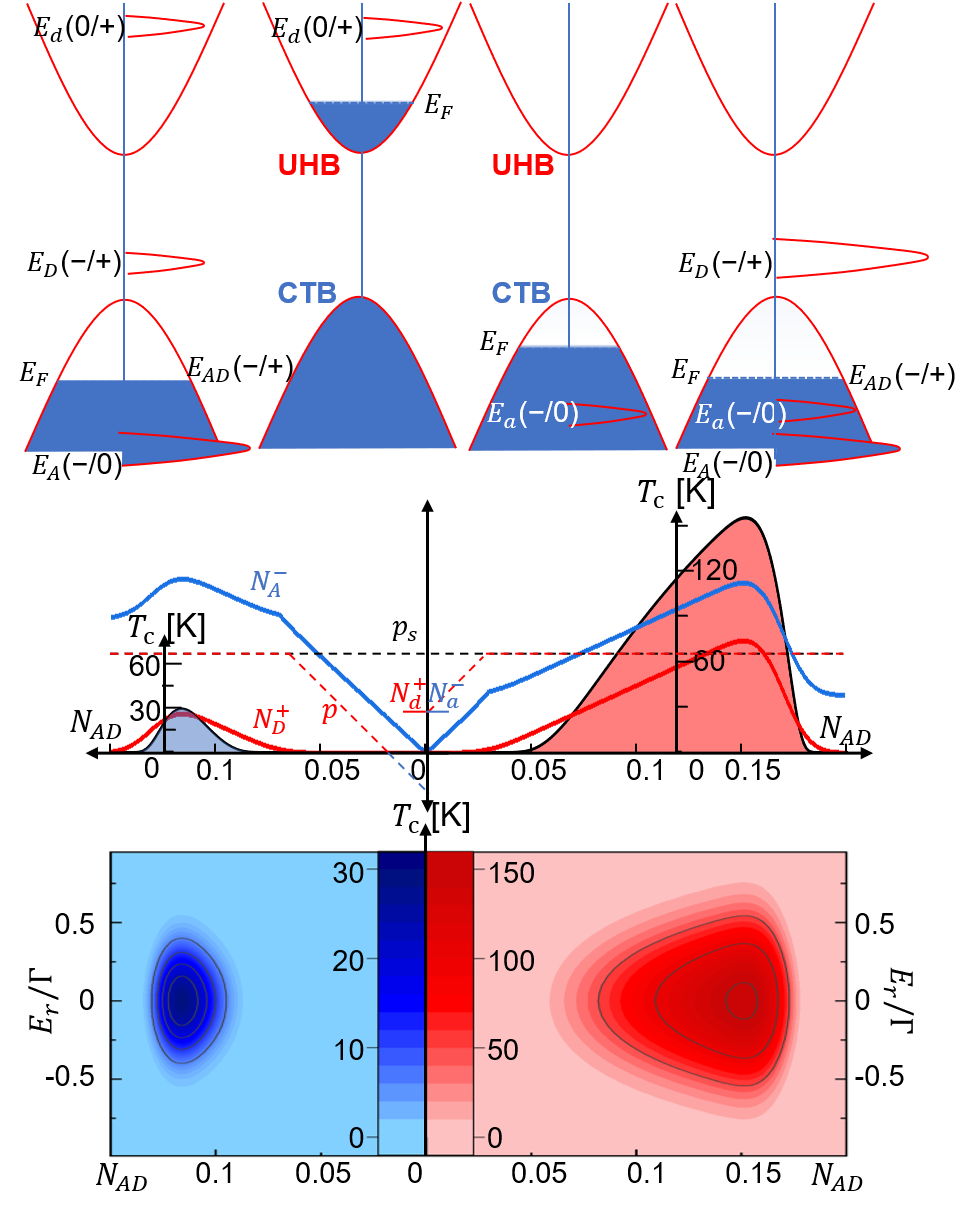}
\caption {Dependence of the critical temperature
$T_c$ on the concentration of amphoteric defects
(\(N_{\text{AD}}\)) for a cuprate chemically doped with \(N_{a}\)
acceptors (right side panel) or \(N_{d}\) donors (left side panel)
assuming the same doping level \(N_{a} = N_{d} = {0.4p}_{s}\). The
calculated $T_c$ was broadened with the same
\(N_{A^{-}}^{\text{cr}} = {2p}_{s}\) for both doping conditions. The straight
lines show the concentrations of free carriers as well as concentrations
of the ADs in the donor (\(D^{+}\)) and acceptor (\(A^{-}\))
configuration. The low panel shows color-coded two-dimensional maps of
$T_c$ vs NAD and the Fermi energy relative to the charge
transition
\( E_{r}/\Gamma = E_{F} - [E_{\text{AD}}(-/+)]/\Gamma\).}
\label{fig:figure5}
\end{figure}

Chemical doping provides an insight into the origin of the double dome
structure in$\ce{YBa2CuO_{6 + \delta}}$ 
where two plateaus' one with \(T_{C} \cong 60\ K\) and the other
\(T_{C} \cong 90\ K\) are observed in the critical temperature
dependence on the excess oxygen content, $\delta$.\cite{Putilin1993,Cava1987} The unique characteristic of
$\ce{YBa2CuO_{6 + \delta}}$ is that in
addition to the standard 2-dimensional \ce{CuO2} planes Cu
and O form also 1-dimensional CuO chains. 
Changing the O content
produces dangling bond like defects in both chains and planes. However,
the dangling bonds in the chains act only as external sources of holes
to be transferred to the conducting \ce{CuO2} planes.
Therefore, the role of the chains is similar to the p-type chemical
doping which shifts the saturation of the hole concentration and thus
also the onset for the formation of \(N_{D^{+}}\) to lower
nonstoichiometry, \(\delta\). The experiments indicate that the hole
concentration originating from the chains reaches the 60 K plateaus for
\(0.7 \geq \delta \geq 0.6\) (\(0.12 \geq N_{\text{AD}} \geq 0.1\)) . At
larger \(\delta\) the concentration of \(D^{+}\) is controlled by holes
originating from the planes leading to the higher $T_c$ of
90 K.\cite{Cava1987a}

In the above considerations we have shown that the superconductivity in
cuprates is associated with the presence of ADs whose concentration
depends on the Cu/O ratio that can be controlled by proper material
processing conditions. It is not overstatement to claim that sample
preparation conditions are the most critical factor determining
superconducting properties of all cuprate superconductors. This is
strongly supported by the fact that the optimum superconducting
properties are frequently achieved with materials processing procedures
based on a trial-and-error approach. Also it was shown that annealing under O-rich or O-deficient
conditions can reversibly change the same material between
superconductor and normal conductor phase.\cite{J.M.TarasconL.H.Greene1987} Such behavior is fully
consistent with our model as these processing conditions provide the
most straightforward way to change the chemical potential of O and thus
also the formation energy and concentration of ADs. Also, it explains
why achieving superconductivity in chemically n-type doped cuprates
always require so extreme processing conditions. It is because in this
case compensation of the chemical donors needed to move the Fermi energy
into the CTB requires more ADs in acceptor configuration that have
detrimental effect on superconductivity.

Although changing material stoichiometry is the most straightforward way
to control concentration of ADs it should be emphasized that at high
temperatures even a stoichiometric material can have high thermodynamic
equilibrium concentration of ADs in the form of O and Cu vacancies. The
defects transform between donor and acceptor configuration and reach the
minimum energy for the Fermi energy stabilized at the charge transition
state, \(E_{\text{AD}}(+/-)\). The high
temperature defect configuration can be frozen with a rapid cooling that
is frequently used in preparation of superconducting cuprates. This
means that, in general, the electrical properties of cuprates are not
determined by chemical doping and/or excess or deficiency of oxygen and
that with proper high temperature processing superconductivity could be
achieved in undoped stoichiometric cuprates. Indeed, extensive annealing
studies aimed at control of the O content have shown that
La\textsubscript{2}CuO\textsubscript{4} which is a parent compound for a
number of p-type doped superconductors\cite{Grant1987}  and
Pr\textsubscript{2}CuO\textsubscript{4} a parent compound n-type
superconductor \cite{Brinkmann1995} exhibit superconducting properties
without or with insignificantly small chemical doping.

All the above considerations lead to very important conclusions
regarding superconducting properties of cuprates. They demonstrate that
amphoteric defects are exclusively responsible for the superconductivity
of mobile holes interacting with hole pairs localized on amphoteric
defects in the donor configuration, \(D^{+}\). The p-type or n-type
chemical doping is neither necessary nor sufficient condition for
superconductivity. The role of chemical doping is limited to changing
the balance between ADs in the donor and acceptor configuration with
p-type (n-type) doping increasing (decreasing) the concentration of
localized \(D^{+}\) centers resulting in enhancement (reduction) of
superconducting properties.

\section{Effects of External Perturbations}

The present calculations of the superconducting properties tacitly
assume that the hole concentration as well as concentrations of ADs in
the donor and acceptor configurations are determined by the low
temperature thermodynamic equilibrium with the Fermi energy in resonance
with \(E_{\text{AD}}(-/+)\). In general, the assumption
cannot be justified as preparation of superconducting cuprates requires
high temperature processing followed by specific cooling conditions
indicating that the low temperature superconducting properties are
determined by the O content and a balance between ADs in donor and
acceptor configuration frozen at some higher temperature. The
temperature has to be high enough to allow for the transformation of ADs
between donor and acceptor configuration but also low enough to assure
that the frozen configuration is as close as possible to the optimum
conditions for the superconductivity. The complexity of the sample
processing procedures to achieve the optimum superconducting properties
of cuprates can be attributed to three energy scales relevant to this
problem. First, shown in \textbf{Fig. 2}(c) energy barrier
E\textsubscript{b} which affects the balance between ADs in the donor
and acceptor configuration second, the negative-$U$ energy and third the
formation energy of the ADs, \(Ef_{A}(E_{F}) = 
Ef_{D}(E_{F})\) whose value for
\(E_{F} = E_{\text{AD}}(-/+)\) determines their total
concentration. The latter energy depends on the non-stoichiometry given
by Cu to O ratio. Additional factor adding to the complexity of the
system is that, as is shown in \textbf{Fig. 5} the concentration of the
ADs in the donor and acceptor configuration can be affected by the
chemical doping.

As has been shown in sections III and IV the highest $T_c$
is obtained for the Fermi energy at the resonance with
\(E_{\text{AD}}(-/+)\) (\(E_{\text{r}} = 0\)) and
with maximized concentration of isolated \(D^{+}\) centers.
Unfortunately, considering the complex nature of the cuprate materials
it is not likely there could be a material processing path leading to
achieving these two goals simultaneously. Consequently, all the
materials processing procedures aimed at maximizing superconducting
properties are compromises between attempts to reach the resonance
conditions while maximizing concentration of isolated \(D^{+}\) defects.
To gain an insight into this intricate process we simplify this problem
by considering two limiting cases. In the first case we assume a
constant concentration of \(D^{+}\) for both p-type and n-type cuprates.
Then, as is shown in \textbf{Fig. 6} (a) initially the Fermi energy is
located below (above) \(E_{\text{AD}}\left( - / + \right)\)for p-type
(n-type) material. An optimization process reduces (increases)
\(N_{A^{-}}\) shifting the Fermi energy towards
\(E_{\text{AD}}\left( - / + \right)\) in both cases. In the optimized
material \(E_{\text{r\ }} \cong 0\) and $T_c$ reaches the
maximum value. In the second limiting case the optimization process aims
at maximizing \(N_{D^{+}}\) even if it is done at the expense of missing
the resonance conditions. As is shown in \textbf{Fig. 6} (e) this leads
to an optimum $T_c = T$\textsubscript{C,op} at
\(E_{\text{r\ }} = E_{\text{r,op\ }} \neq 0\). It is obvious that those
two limiting cases are gross simplifications of much more complex
processes. However, they provide intuitive insights for understanding of
the mechanism of the optimal doping as well as effects of external
perturbation on the superconducting properties of cuprates.

Results in \textbf{Figure 4} show that the first of the discussed above
limiting cases applies to the doping region where the ADs are well
isolated and do not interact with each other and the maximized
$T_c$ is solely determined by the concentration of
\(N_{D^{+}}\) corresponding to the resonance condition
(\(E_{\text{r\ }} \cong 0\)). This leads to the experimentally observed
universal dependence of $T_c$ on the superfluid density in
the underdoped region.\cite{Uemura1989} However, increasing the
concentration of ADs can change the optimum processing conditions
towards the second limiting case where an increase in $T_c$
associated with increase of the superfluid density \(N_{D^{+}}\) is
compensated by reduction of $T_c$ caused by shift of the
Fermi energy to off resonance conditions (\(E_{\text{r\ }} \neq 0\)).
This results in a saturation or even a decrease of $T_c$
despite an increase of the superfluid density \(N_{D^{+}}\).\cite{Uemura1989,Niedermayer1993} 
Such material specific saturation of the
optimum $T_c$ is clearly observed in superconducting
cuprates.\cite{Uemura1989} As has been discussed in section III at
even higher concentration of ADs corresponding to the overdoped region
the superfluid density is reduced as \(D^{+}\) are passivated by
\(A^{-}\). Again, $T_c$ becomes proportional to
\(N_{D^{+}}\) until it is terminated when all \(D^{+}\) are passivated.
This explains experimentally observed linear relationship between
$T_c$ and superfluid density found in the overdoped
region.\cite{Niedermayer1993}

An interesting and perplexing feature of superconducting cuprates is
their unusually complex dependence of $T_c$ on hydrostatic
pressure.\cite{Markert1990,Griessen1987} A large number of studies have shown
that pressure dependencies of the critical temperature can vary widely,
with $T_c$ pressure coefficients ranging from large
positive to large negative values. More systematic studies have
identified some important trends. Thus, it was found that application of
hydrostatic pressure increases (decreases) $T_c$ in p-type
(n-type) doped cuprates with the pressure induced $T_c$
change larger for non-optimally doped than for optimally doped
materials.\cite{Markert1990} Furthermore, it was found that in some
p-type cuprates optimized for the maximum $T_c$ application
of very higher pressures results in a nonmonotonic dependence with an
initial rapid increase followed by saturation or even decrease of
$T_c$ at the highest pressures.\cite{Gao1994} Here we
will show that the main characteristic features of these distinct
pressure effects on $T_c$ can be understood in terms of the
discussed above thermodynamic properties of ADs.

In order to evaluate the pressure effects on $T_c$ one
needs to know how the pressure affects the charge transfer band (CTB)
edge energy. Previous studies on charge transfer oxides have shown that
application of hydrostatic pressure increases the width of the CTB and
reduces the charge transfer gap.\cite{Obradors1993} Adopting these
findings and considering that the charge transition state
\(E_{\text{AD}}\left( - / + \right)\) of highly localized ADs is
independent of pressure\cite{Nolte1987}  indicates that application
of hydrostatic pressure will result in an upward shift of the CTB edge
and thus also the Fermi energy relative to
\(E_{\text{AD}}\left( - / + \right)\) directly affecting the resonance
conditions for the maximum $T_c$. This leads to a distinct
asymmetry in the effects of pressure on the $T_c$ in p-type
and n-type superconducting cuprates. Specifically, as is shown in the
left side top panel of \textbf{Fig. 6} the pressure induced upward shift
of the CBT edge moves the Fermi energy towards (away) from the resonance
resulting in an increase (decrease) of $T_c$ for p-type
(n-type) cuprates. The initial relative pressure coefficient of
$T_c$ ($d\ln T_c/dP$) is determined by shown in
\textbf{Fig. 6} bell-like shape of the dependence of $T_c$
on the deviation from the resonance condition, \(E_{\text{r\ }} = 0\).
\textbf{Figure 6} (h) shows that the calculated pressure coefficient
changes from large positive values for non-optimal p-type doping goes to
zero for optimal doping and becomes negative for non-optimal n-type
doping. This is in a good qualitative agreement with experimentally
observed trends.\cite{J.M.TarasconL.H.Greene1987} Here $T_c$ was
calculated with \(N_{D^{+}} = 0.04\ \)and the other parameters the same
as those used for the results shown in \textbf{Figs. 3} and \textbf{4}.
The only additional parameter is the pressure induced shift of the Fermi
energy relative to the charge transition state i.e.
\( dE_{r}/dP = d(E_{F} - E_{\text{AD}}(-/+))/dp\)
which is equal to the pressure induced upward shift of the CTB edge
energy. Here we find that an overall good explanation of the
experimentally observed trends is obtained for
\( d(E_{r}/\Gamma)/dp = 0.01/\)GPa or
\( dE_{r}/dP =\) 2 meV/GPa which is in the range of
typical values of pressure induced shifts of the valence band edges in
semiconductors.\cite{Nolte1987}

\textbf{Figures 6} (d), (e) and (f) illustrate the effects of
hydrostatic pressure on $T_c$ for the second discussed
above limiting case where the optimum $T_c =
T_{c,\textrm{op}}$ is achieved with maximized \(N_{D^{+}}\). In order
to model the pressure effects in
$\ce{HbBa2Ca_{m-1}O_{m+2+\delta}}$
 with $m = 1,2,3$ it is assumed that the optimized
\(N_{D^{+}}\) is increasing with $m$.\cite{Gao1994}
Specifically, the calculations shown in \textbf{Fig. 6} (i) and (j) were carried out for three different
concentrations \(N_{D^{+}} =\) 0.035, 0.04 and 0.045 with the assumption
that in all instances the optimal $T_c$ is reached for
\( E_{r}/ \Gamma = - 0.3\). Again, the other parameters have the
same values as used in the previous calculations. It is seen that the
pressure induced upward shift of the CTB edge moves the Fermi energy
towards the resonance with \(E_{\text{AD}}(-/+)\)
resulting in an increase of $T_c$. The $T_c$
reaches the maximum T\textsubscript{Cr} for \(E_{r} = 0\ \) and then
decreases as \(E_{r}\) becomes larger than 0. These calculations are in
a very good qualitative agreement with experiments.\cite{Gao1994}
They show that the optimized $T_c$ at the ambient pressure
and the maximum $T_c$ at about 30 GPa exhibit a strong
dependence on \(N_{D^{+}}\) and thus also on the number of
\ce{CuO2} planes, m in the unit cell. In contrast the
pressure induced changes of
\(T_{C}\left( P \right) - T_{C}\left( P = 0 \right)\) show very similar
almost universal dependencies for all values of \(N_{D^{+}}\). This
result is a simple reflection of the fact that the \(N_{D^{+}}\) has a
negligible effect the bell-shaped dependence of $T_c$ on
\(N_{D^{+}}\).

\begin{figure}
\centering
\includegraphics[width=5.96364in,height=2.75465in]{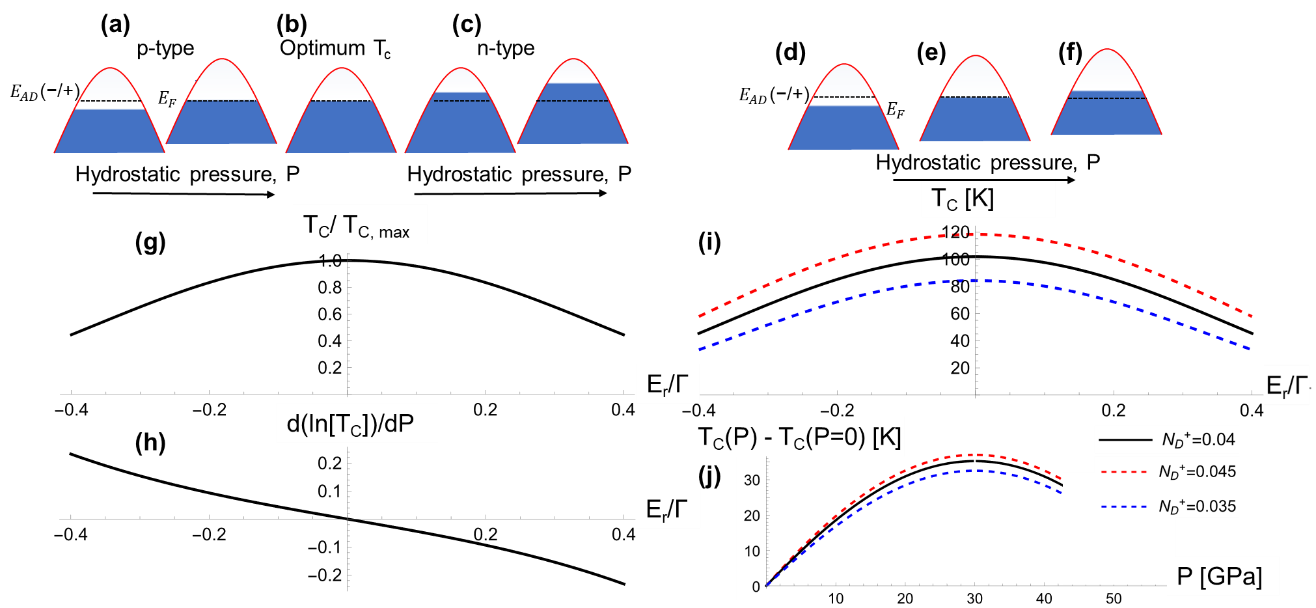}

\caption{(\textbf{Left panel, a, b, c, g, h}) Dependence of
$T_c$ on hydrostatic pressure for p-type and n-type
cuprates. In the p-type application of hydrostatic pressure to
non-optimally doped cuprate (\(E_{r} < 0\)) shifts CTB and the Fermi
energy towards resonance resulting in positive pressure coefficient. In
non-optimally doped n-type cuprate (\(E_{r} > 0\)) application of
pressure shifts the Fermi energy away from the resonance resulting in a
negative pressure coefficient. The pressure coefficient for optimal
doping (\(E_{r} = 0\)) equals 0 at low pressures. (\textbf{Right panel,
d, e, f, i, j}) Pressure dependence of $T_c$ optimized with
maximum \(N_{D^{+}}\) and \(E_{r} < 0\). The upward shift of CTB and the
Fermi energy increases $T_c$ until it reaches a maximum
value \(E_{r} = 0\). Further increase of the pressure leads to a decrease
in $T_c$ as the Fermi energy shifts to off resonance
conditions \(E_{r} > 0\).}
\label{fig:figure6}
\end{figure}

Another characteristic and unusual feature of superconducting cuprates
is a strong dependence of the critical temperature on the external bias\cite{Bollinger2011,Leng2011} where it was shown that application of high
electric field can vary properties of thin cuprate films between
superconducting and insulating state. This is easy to understand in our
model as the applied field affects the hole concentration and shifts the
Fermi energy relative to the charge transition state
\(E_{\text{AD}}(-/+) \) affecting the resonance
condition and thus also critical temperature. It is important to note
that such external bias induced transition can be observed only for
material with \(N_{D^{+}} > 0\), that is, \(N_{\text{AD}} > p_{s}\).

\section{Summary and Outlook}

In this paper we have presented a model of superconductivity mediated by
amphoteric, negative-$U$ defects. The model is based on the vacancy-like
defects in \ce{CuO2} planes that can undergo a transformation
between acceptor and donor configuration stabilizing Fermi energy and
the hole concentration in the charge transition band. The amphoteric
defects in the donor configuration trap two holes and act as highly
localized preformed hole pairs facilitating the superconducting
transition of the free hole gas. The critical temperature is fully
determined by the concentration of isolated amphoteric defects in the
donor configuration and deviation of the Fermi energy from the energy of
the charge transition state between acceptor and donor configuration.
The model provides a unified explanation for key properties of the
superconducting cuprates listed in points 1) to 5) in the introduction.

\begin{enumerate}
\def\labelenumi{\arabic{enumi})}
\item
  It is shown that the superconductivity in cuprates is innately related
  to presence of amphoteric defects in \ce{CuO2} planes.
  Elaborate sample processing conditions with controlled Cu/O content
  ratio are required to maximize the concentration of preformed hole
  pairs \(N_{D^{+}}\) while maintain the charge balance stabilizing the
  Fermi energy close to the resonance with the ADs charge transition
  state. (See \textbf{Figs. 3, 5} and \textbf{6})
\item
  The dome-like dependence of the critical temperature is explained by
  doping induced increase of the concentration of performed hole pairs
  in the underdoped region. The reduction and the termination of $T_c$ at
  high doping levels is associated with passivation of the amphoteric
  defects in the donor and acceptor configuration (see \textbf{Fig. 4}).
  More subtle effects come into play close to the optimum doping where
  the critical temperature is affected by both the superfluid density
  and the deviation from resonance conditions. (See \textbf{Figs. 3} and
  \textbf{6})
\item
  The universal dependence of the $T_c$ on the superconducting phase is the
  reflection of the fact that for the resonant coupling between free
  holes and preformed hole pairs the critical temperature is uniquely
  and in the same way affected by the superfluid density given by the
  concentration of the amphoteric defects in the donor configuration in
  all cuprates. (See \textbf{Fig. 4})
\item
  It is shown that chemical doping with acceptors or donors is neither
  sufficient nor necessary condition for superconductivity which in both
  cases is associated with coupling of free holes in the charge transfer
  band to hole pairs localized on amphoteric defect donors. The
  asymmetry in the Tc dependence on the doping originates from higher
  (lower) concentration of the defect donors in p-type (n-type) doped
  cuprates. (See \textbf{Fig. 5})
\item
  The complex trends in the hydrostatic pressure dependence of the
  critical temperature are explained by the change in the resonance
  conditions resulting from the upward pressure induced shift of the
  charge transfer band and thus also the Fermi relative to the charge
  transition state of the ADs. (See \textbf{Fig. 6})
\end{enumerate}

This paper is focused on the effects of ADs on the superconducting
properties of cuprates. However, it is obvious that the amphoteric
defects play also critical role in determining normal state properties
of these materials. In particular the electrical properties of the
normal state are determined by the transport of free holes in the charge
transfer band scattered by charged ADs in the donor and acceptor
configuration. Furthermore, at higher temperatures electric charges can
be transported within a defect band formed at the charge transition
state \(E_{\text{AD}}(-/+)\). The complex conductivity
mechanism appears to be consistent with the observation that
superconducting cuprates do not exhibit typical metallic, Fermi liquid
like normal state properties.\cite{Norman2003} Another important
issue is the extent to which the properties of normal state cuprates are
affected by large concentration of the ADs in the acceptor configuration
that is needed to stabilize the Fermi energy. Of particular significance
could be the effect of resonance scattering of hole pairs on free of
holes \(A^{-}\) centers. This scattering is expected to affect the
density of states in the charge transfer band and could be responsible
for the normal state pseudogap effect that is routinely observed in
superconducting cuprates.\cite{Norman2005} These and other issues
relating to the normal state of superconducting cuprates will be
addressed in future work.

The present paper is entirely devoted to the high $T_c$ cuprates that are
by far the most prominent and representative group of unconventional
superconductors in which there is convincing evidence that the
superconductivity is not mediated by the electron phonon interaction.
However, the class of unconventional superconductors is quite broad and
includes a large variety of materials with varying degree of chemical
complexity.\cite{Stewart2017} Interestingly properties of some of
these superconductors show a similarity to the properties of cuprates.
Thus, the broad class of iron based superconductors (IBSC)\cite{Hosono2015}
 with critical temperature of more than 50 K and
semiconducting SrTiO\textsubscript{3} perovskite with $T_c$ of less than 1
K show a well-developed dome-like dependence of
the critical temperature on various forms of doping.\cite{Bustarret2015,Schooley1964} Again, it was shown
that in both materials systems the superconductivity can be produced by
varying only the crystal nonstoichiometry without any intentional
chemical doping.\cite{Hosono2015,Schooley1965} Furthermore a large electron
accumulation on the free surface of SrTiO\textsubscript{3}\cite{DeSouza2014}
and at the superconducting interface between
\ce{SrTiO3}/LaAlO\textsubscript{3} \cite{Thiel2006} provide a
strong evidence for an existence of localized defect charge transition
state in the conduction band of SrTiO\textsubscript{3}. All these
observations strongly suggest that it could be interesting to look at
the possibility that negative-$U$ amphoteric defects are also responsible
for superconductivity in these and other similar materials.


\section*{Author contributions}
\noindent
\hl{Conceptualization: WW \\
Formal analysis: SW, WW \\
Funding acquisition: JWA, WW \\
Methodology: SW, WW\\
Software: SW \\
Supervision: WW \\
Validation and Visualization: SW \\
Writing - original draft: SW, WW \\
Writing - review and editing: SW, JWA, WW}

\begin{acknowledgments}
\hl{This work was partially funded by the National Science Centre (Poland) NCN Opus grant no. UMO-2019/33/B/ST3/03021. Initial work was supported by by the Singapore-Berkeley Research Initiative in Sustainable Energy (SinBeRISE) which is supported by the National Research Foundation (NRF) of Singapore. JWA was supported by the U.S. Department of Energy, Office of Science, Office of Basic Energy Sciences, Materials Sciences and Engineering Division under contract no. DE-AC02-05CH11231 within the Electronic Materials Program (KC1201).}
\end{acknowledgments}

\section*{Data Availability Statement}

\hl{The data that support the findings of this study are available upon reasonable request from the corresponding authors. }

\bibliography{HighTc}
\end{document}